# Mechanical Dissipation in Silicon Flexures

S. Reid*, G. Cagnoli, D.R.M. Crooks, J. Hough, P. Murray, S. Rowan
*University of Glasgow, UK*

M.M. Fejer, R. Route, S. Zappe
*Stanford University, USA*

The thermo-mechanical properties of silicon make it of significant interest as a possible material for mirror substrates and suspension elements for future long-baseline gravitational wave detectors. The mechanical dissipation in 92μm thick <110> single-crystal silicon cantilevers has been observed over the temperature range 85 K to 300 K, with dissipation approaching levels down to $\phi = 4.4 \times 10^{-7}$.

*email: s.reid@physics.gla.ac.uk

## I. INTRODUCTION

Long baseline gravitational wave detectors operate using laser interferometry to sense the differential strain, caused by the passage of gravitational waves, between mirrors suspended as pendulums. These detectors operate over a frequency range between the pendulum modes of the suspensions (typically few Hz) and the lowest internal resonances of the mirrors (few 10's of kHz). One important limit to the displacement sensitivity of current and planned detectors in the frequency range of operation is off-resonace thermal noise in the mirrors and suspensions driven by thermal fluctuations. Thus low mechanical loss materials, such as silica, sapphire and silicon are currently used or proposed for detectors at the forefront of this research.

Improved sensitivity at low frequencies (few Hz to few 100 Hz) will require further reduction in the level of thermal noise from the test masses and their suspensions. A possible route for achieving this is through cooling. Fused silica, the most commonly used test mass material, exhibits a broad dissipation peak at around 40 K and therefore is not a promising candidate for cooling [1]. Sapphire and silicon however are good candidates. Work is currently being carried out in Japan on developing cooled sapphire test masses and suspension fibers for use in a transmissive Fabry-Perot based interferometer [2,3], and in Europe and the US research is underway on the use of silicon at low temperatures [4,5].

At higher frequencies (greater than a few 100 Hz) the performance of current interferometers is not limited by thermal noise from the optics but by photo-electron shot noise, whose significance can be reduced by circulating higher optical powers in the interferometer. However, power absorbed by the test masses and mirror coatings can cause excessive thermally induced deformations of the optics, causing the interferometer to become unstable. The extent of this deformation is proportional to $\alpha/\kappa$ [6], where $\alpha$ is the linear coefficient of thermal expansion and $\kappa$ the thermal conductivity of the test mass material. Changing from a transmissive to a reflective topology could eliminate thermal loading from substrate absorption, provided coatings of suitably low transmission are available. Used in such a topology, the high thermal conductivity of a silicon mirror substrate would allow circulating powers approximately seven times higher than could be supported by sapphire for the same induced surface deformation making silicon of significant interest as a test mass substrate from a thermal loading standpoint [4].

At room temperature the thermal noise resulting from thermo-elastic effects in interferometers using crystalline optics has been predicted to be a significant noise source in the frequency band of gravitational wave detection [7]. The level of intrinsic and expected thermo-elastic dissipation in silicon is broadly comparable to sapphire at room temperature [8]. However on cooling the linear thermal expansion coefficient of silicon becomes zero at two temperatures, ~125 K and ~18 K [9], and thus around these two temperatures thermoelastic dissipation could be expected to be negligible. It is thus of interest to study the temperature dependence of mechanical dissipation in silicon samples for potential use as suspension elements and mirror blanks. This paper is restricted to studies of thin silicon flexures.

Studies of dissipation in silicon samples of a variety of geometries and types have been carried out by other authors. In particular, dissipation in silicon flexures has been studied in samples of the type used in Atomic Force Microscopes [10,11]. However these cantilevers have dimensions considerably smaller than would be suitable for use in the test mass suspensions of gravitational wave interferometers and thus are in a regime where measured dissipation may be dominated by different sources of dissipation than the dimensions that have been studied here [12].

## II. EXPERIMENTAL PROCEDURE

The single-crystal cantilevers tested were fabricated from a silicon wafer by a hydroxide chemical etch. The anisotropic nature of such etching allows the reduction of thickness whilst a masked, thick end can remain as a clamping block to reduce any 'slip-stick' losses as the cantilever flexes [13]. The geometry of the cantilevers obtained is shown in Fig. 1. The silicon was boron doped with a resistivity of 10-20 Ωcm.

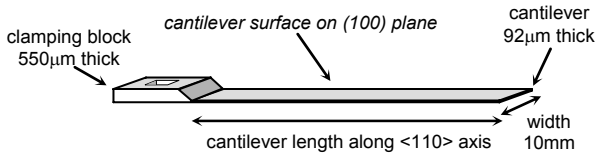

Fig. 1. Schematic diagram of one silicon cantilever tested.

The thick end of each cantilever was held in a stainless steel clamp and placed within a cryostat, shown in Fig. 2, evacuated to approximately $3 \times 10^{-6}$ mb. The resonant modes of each cantilever were excited in turn using an electrostatic drive plate. Laser light reflected from the silicon surface and directed onto a photodiode external to the cryostat allowed the angular motion of the end of the cantilever to be detected. The length of the lever arm due to the optical pipe leading to the inner experimental chamber of the cryostat made the readout system very sensitive to the cantilever motion. As a consequence, loss measurements on the first bending mode were not possible since the readout system saturated before the mode was excited to a level significantly above the background excitation due to ground vibrations. However it was possible to measure the frequency of the first resonance, and this is used later in section II.

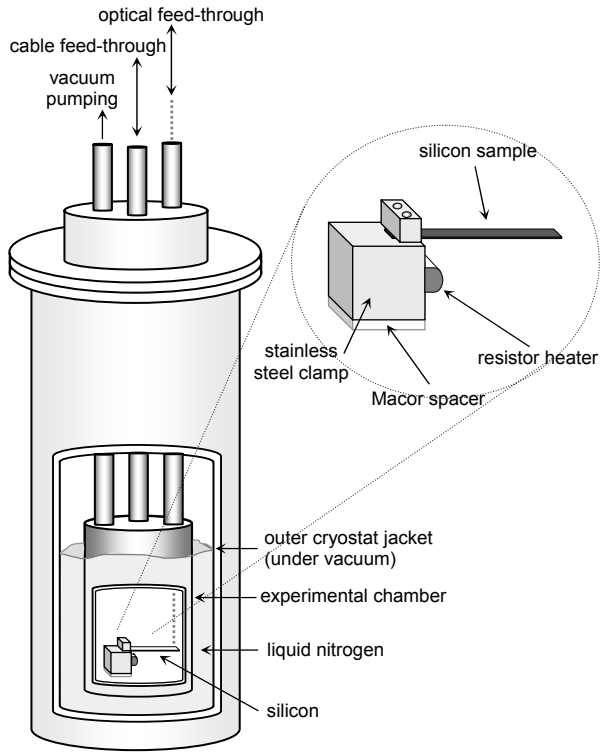

Fig. 2. Schematic diagram of the cryostat.

The mechanical quality factor $Q$ of a resonance of angular frequency $\omega_0$ can be calculated from measurements of the amplitude $A$ of freely decaying resonant motion. It can be shown that the time dependence of the amplitude decay is given by

$$A = A_0 e^{-\omega_0 t/(2Q)}, \qquad (1)$$

where $A_0$ is the initial amplitude of the motion. The mechanical loss $\phi(\omega_0)$ is the inverse of the quality factor [14,15]. The mechanical losses of several modes of each cantilever were measured at temperatures from 85 K to 300 K. Presented here are loss measurements for the third (f ~ 670 Hz) and fifth (f ~ 2185 Hz) bending modes of a cantilever 57 mm long and the third mode (f ~ 1935 Hz) of a shorter 34 mm long cantilever.

## A. Temperature dependence of mode frequencies

At a given frequency the expected thermoelastic dissipation depends on the sample thickness. The thickness of the silicon samples was measured to be $(92 \pm 2)$ μm using a Wyko NT1100 Optical Profiler.

The frequency of each bending mode changes as the silicon is cooled due to the temperature dependence of Young's modulus, $E(T)$. $E(T)$ can be calculated using the semi-empirical formula [16],

$$E(T) = E_0 - BT \exp\left(-\frac{T_0}{T}\right), \qquad (3)$$

where $E_0$ is the Young's Modulus at 0 K, $B$ is a temperature independent constant related to the bulk modulus, $T$ is the temperature in Kelvin and $T_0$ is related to the Debye Temperature.

The angular resonant frequency $\omega$ of the third bending mode of a homogeneous beam of thickness $t$ and length $L$ is given by [17],

$$\omega = (7.853)^2 \frac{t}{L^2} \sqrt{\frac{E}{12\rho}}, \qquad (4)$$

where $\rho$ is the material density. As noted by Gysin et al. [18] any change in $\omega$ resulting from a temperature dependant variation in $T$, $L$ or $\rho$ is smaller than that from the variation in $E$ and may be ignored.

Using the value $T_0 = 317$ K [18], Eqs. (3) and (4) were used to find a best-fit curve to the observed temperature dependence of the frequency and thus a value for $E_0$ obtained. It was possible to measure the first cantilever length to be 57.0 mm to an accuracy of ±0.5 mm without contacting the cantilever surface. The second cantilever length was found to be $(34.0 \pm 0.5)$ mm.

For the first cantilever, the third mode saw the best agreement between the predicted and experimental frequencies with an $E_0$ value of $(161.7 \pm 0.8)$ GPa. The temperature dependence of the calculated and measured frequencies for this mode are shown in Fig. 3. Applying the same model to the first and fifth resonant modes of this sample (approximately 39 Hz and 2185 Hz) gave very similar values for $E_0$. The average value of $E_0$ with associated standard error is $(163 \pm 4)$ GPa.

Likewise, the third and fifth bending modes for the cantilever of length 34 mm matched the predicted frequencies when $E_0 = (165 \pm 6)$ GPa. Combining the calculated $E_0$ values yields $(164 \pm 3)$ GPa which appears close to the literature value $E_0 = 167.5$ GPa [19].

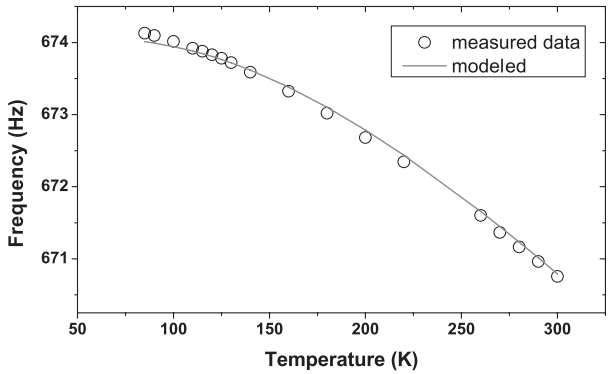

Fig. 3. Temperature dependence of the frequency of the third resonant mode at f ~ 670 Hz of the cantilever of length 57 mm.

## III.  LOSS AS A FUNCTION OF TEMPERATURE

The measured loss, $\phi_{\text{measured}}(\omega)$, is the sum of dissipation arising from a number of sources,

$$\phi_{\text{measured}}(\omega) = \phi_{\text{thermoelastic}}(\omega) + \phi_{\text{bulk}}(\omega) + \phi_{\text{surface}}(\omega)$$
$$+ \phi_{\text{clamp}}(\omega) + \phi_{\text{gas}}(\omega) + \phi_{\text{other}}(\omega), \quad (4)$$

where $\phi_{\text{thermoelastic}}(\omega)$ is loss resulting from thermoelastic damping, $\phi_{\text{bulk}}(\omega)$ is the bulk (or volume) loss of the material, $\phi_{\text{surface}}(\omega)$ is the loss associated with the surface layer, $\phi_{\text{clamp}}(\omega)$ is the loss associated with the clamping structure, $\phi_{\text{gas}}(\omega)$ is the loss due to damping from residual gas molecules and $\phi_{\text{other}}(\omega)$ is loss from any other possible dissipation process.  In order to estimate the level of thermal noise expected from using silicon in gravitational wave detectors test masses and suspensions, $\phi_{\text{thermoelastic}}(\omega)$, $\phi_{\text{bulk}}(\omega)$ and $\phi_{\text{surface}}(\omega)$ must be quantified. Therefore in our experiment all the other sources of loss must be minimised.

### A.  Thermoelastic loss

Thermoelastic loss is associated with the flexing of a thin suspension element where the cyclical stretching and compression of alternate sides of a flexing sample results in heat flow between the compressed and expanded regions [20].  The flow of heat is a source of loss.  In the simple case of a bending bar of rectangular cross section, the thermoelastic loss can be expressed as,

$$\phi_{\text{thermoelastic}}(\omega) = \Delta \frac{\omega\tau}{1+\omega^2\tau^2}, \quad (5)$$

where $\Delta = \dfrac{E\alpha^2 T}{\rho C}$ (6)

and $\tau = \dfrac{1}{\pi^2}\dfrac{\rho C t^2}{\kappa}$ (7)

with $\tau$ the characteristic time for heat transfer across the bar, $C$ is the specific heat capacity of the material and other parameters are as defined earlier.

Eqs. (5) to (7) may be used to calculate the temperature dependent thermoelastic loss for our sample using the relevant material parameters.  Table 1 shows the room temperature parameters used.  The temperature dependent parameters were taken from 'Thermophysical Properties of Matter' (Touloukian) [9], except for the value of Young's Modulus (E) which was calculated using Eq. (3).  Data for the coefficient of thermal expansion comes from the recommended curve, Vol. 13 p.155, and the specific heat from curve 2, Vol. 5 p.204. The thermal conductivity data is taken from the curves presented in Vol. 2 p.326.  Here minimum, median and maximum values are taken to represent the spread of data for single-crystal silicon at each temperature.

Table 1
Room temperature parameters for silicon [9].

| Parameter | Magnitude | |
|---|---|---|
| Young's modulus (E) | 162.4 GPa | |
| Coefficient of linear thermal of expansion ($\alpha$) | $2.54\times10^{-6}$ K$^{-1}$ | |
| Density ($\rho$) | 2330 kg m$^{-3}$ | |
| Specific heat capacity (C) | 711 J kg$^{-1}$ K$^{-1}$ | |
| Thermal conductivity ($\kappa$) | min:    130<br>median: 145<br>max:    160 | } W m$^{-1}$ K$^{-1}$ |

The uncertainty in the calculated magnitude of the thermoelastic loss comes predominantly from this variation in thermal conductivity ($\kappa$) between silicon samples.

### B.  Results

The measured mechanical losses of the fifth and third bending modes of the silicon cantilever of length 57 mm are shown in Figs. 4 and 5.  The results obtained from the second, forth and sixth bending modes showed similar trends across the temperature range.  The measured mechanical loss of the third bending mode of the silicon cantilever of length 34 mm is shown in Fig. 6. Plotted alongside are the predicted levels of thermoelastic dissipation for each mode.

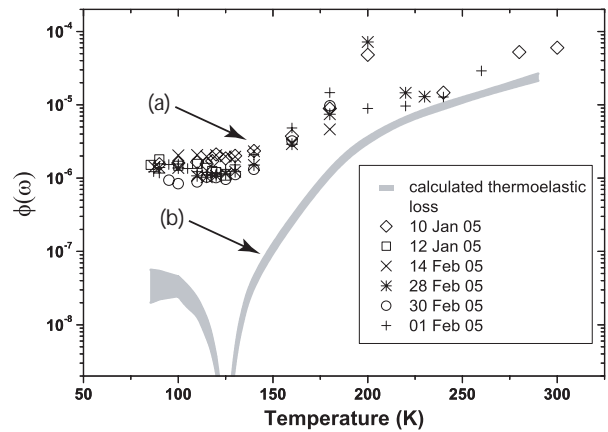

Fig. 4.  Temperature dependence of (a) measured loss, and (b) calculated thermoelastic loss, for the fifth bending mode at 2185 Hz, for cantilever of length 57 mm.

Each data point in Figs. 4 to 6 represents the average of at least three consecutive loss measurements.  To investigate the reproducibility of the measurements, the sample was repeatedly cooled to an initial temperature of ~85 K and loss measurements made as temperature was increased.

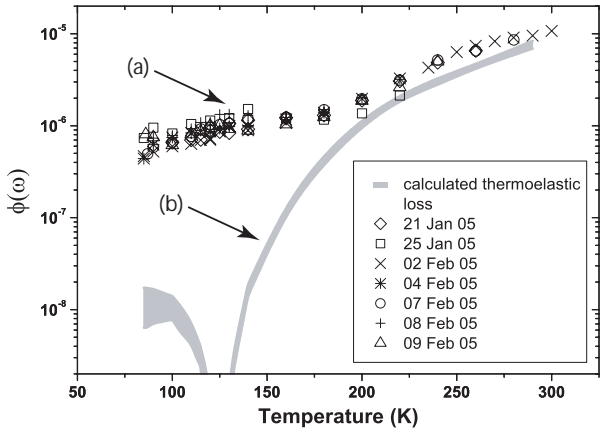

Fig. 5. Temperature dependence of (a) measured loss, and (b) calculated thermoelastic loss for the third bending mode at 670 Hz, for cantilever of length 57 mm.

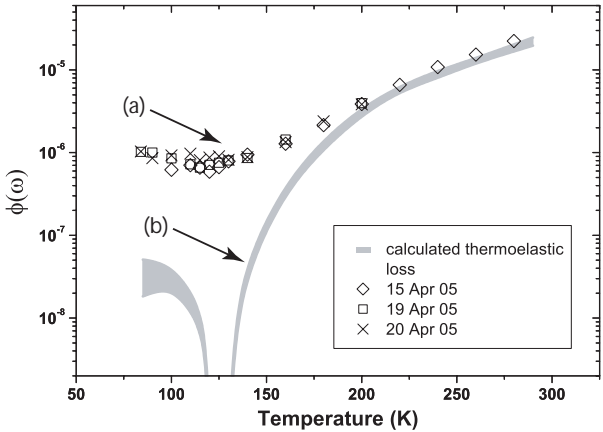

Fig. 6. Temperature dependence of (a) measured loss, and (b) calculated thermoelastic loss for the third bending mode at 1935 Hz, for cantilever of length 34 mm.

The data shows a number of interesting features. First consider the measured mechanical loss of the fifth bending mode as shown in Fig. 4. Measurements of the mechanical loss at temperatures between 85 K and 150 K when made on different days could differ by up to a factor of ~ 2.4.

Also, during two measurement runs a broad dissipation peak was observed at around 200 K. However this is unlikely to result from an intrinsic loss mechanism in the sample since the peak is not observed in the results from 1$^{st}$ February. We believe that these effects are due to energy coupling into the clamping structure. This evidence of a coupling that is dependent on the temperature distribution inside the system, which may differ from run to run, is studied in section IV.

In contrast, the temperature dependence of the loss factors of the third bending modes for both cantilevers, shown in Figs. 5 and 6, showed no sign of any dissipation peaks and appeared to have a smaller variation between experimental runs. It can be seen that the dominant loss mechanism at temperatures above approximately 160 K is broadly consistent with thermoelastic effects. Possible candidates for the additional sources of loss observed at lower temperatures are discussed in the following section.

## IV. DISCUSSION

### A. Surface Loss

Mechanical loss measurements carried out on silicon samples of sub-micron thickness suggest that the measured loss is dominated by surface losses [18,21]. These may be due to a combination of the following:

1. a thin layer of oxidized silicon on the surface [21],
2. shallow damage to the crystal structure (atomic lattice) from surface treatment,
3. contaminants absorbed on or into the surface from the surroundings or from polishing,
4. general (or local) surface roughness [23].

Yasumura et al. measured the loss factors of single-crystal cantilevers with thickness in the range 0.06 to 0.24 μm and found they could be represented by [12],

$$\phi_{surface} = \frac{6\delta}{t} \frac{E_1^S}{E_1} \phi_S,$$  (8)

where $\phi_{surface}$ is the limit to the measurable loss of a cantilever of thickness $t$ and Young's modulus $E_1$ set by the presence of a surface layer of thickness $\delta$, Young's modulus $E_1^S$, and loss $\phi_s$. If for simplicity we assume $E_1 \approx E_1^S$, then Eq. (8) may be used to estimate the limit to measurable dissipation for our sample, set by surface loss, by scaling with thickness the results of Yasumura et al.

The magnitude of this scaled loss, $\phi_{surface}(\omega)$, summed with the upper limit to thermoelastic loss, $\phi_{thermoelastic}(\omega)$, for the third bending mode is shown in Fig. 7. Recall the measured loss varies from run to run and is most likely due to changes in the system during different cycles of cooling and heating. Since this spread is not intrinsic to the sample, the minimum measured losses are presented at each temperature point for comparison. It can be seen that below 160 K the sum of the estimated surface and thermoelastic loss is still lower than the experimental loss by up to a factor of six, thus other loss mechanisms are of a significant level.

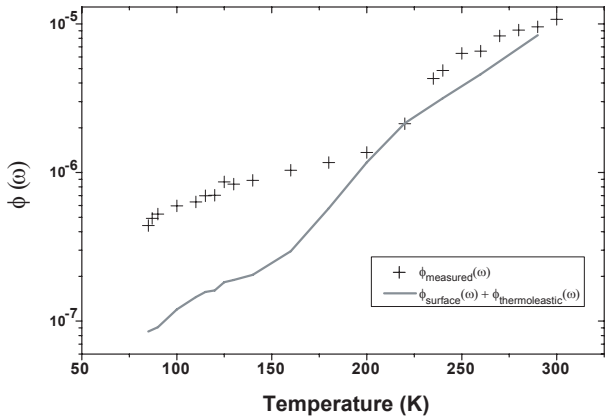

Fig. 7. Plot of (a) the minimum measured loss of the third bending mode at f ≈ 670 Hz for cantilever of length 57 mm compared with (b) the sum of estimated surface loss and calculated thermoelastic loss.

## B. Gas Damping

Suitable vacuum pressures must be reached in order to avoid the measured loss being limited by the result of damping from residual gas molecules in the system. At room temperature the recorded range of gas pressures was $2 \times 10^{-6}$ to $4 \times 10^{-6}$ mb. However, an accurate measure of the gas pressure within the experimental chamber was not possible as the sensor was some distance from the experimental chamber. A residual gas analyser sensitive to molecular weights up to 200 indicated that residual molecules were mainly nitrogen ($N_2$) and water ($H_2O$).

The level of loss due to gas damping of an oscillator can be expressed as [15],

$$\phi_{gas}(\omega) \approx \frac{AP}{m\omega}\sqrt{\frac{M}{RT}},\qquad(9)$$

where $A$ is the surface area, $P$ is the pressure, $m$ is the mass of the oscillator, $\omega$ is the angular frequency of the resonant mode, $M$ is the mass of one mole of the gas, $R$ is the gas constant and $T$ the temperature. For gas damping to be the dominant loss mechanism at room temperature the gas pressure in the experimental chamber would be approximately $2.6 \times 10^{-2}$ mb, assuming the gas to be $N_2$, which is much higher than could be reasonably expected.

The residual gas pressure is expected to decrease with temperature. In Figs. 4 and 5, below 160 K the level of dissipation for both modes at temperatures at best approaches levels around $10^{-6}$. The lack of a clear $1/\omega_0$ dependence here is also consistent with gas damping not being significant.

## C. Bulk Loss

Measurements by us and by other authors have shown that the intrinsic bulk dissipation of single-crystal silicon cylinders at room temperature can be as low as $2 \times 10^{-8}$ to $7 \times 10^{-9}$ and in general is found to decrease as temperature decreases [24-26]. This suggests that the bulk loss of silicon is significantly lower than the measured losses of the cantilever studied here.

However the measurements of McGuigan et al. [25] revealed dissipation peaks near the temperatures where the coefficient of thermal expansion goes to zero. Other experimenters have also observed dissipation peaks at these, and other temperatures, see for example [12].

There are a variety of explanations postulated in the literature for the existence of each of the peaks observed however there appears no reason that the peaks should be related to the zeros in the coefficient of expansion. Therefore it is of interest to investigate whether such peaks in the loss are observed in the sample being studied here. Over the temperature range from 115 to 130 K there is a plateau in the loss in all the modes presented in Figs. 4 to 6. There is no sign of a clear dissipation peak within the temperature range of these measurements, at the levels of dissipation observed.

## D. Clamping Loss

For the case of a two-dimensional system radiating into a semi-infinite silicon substrate the structural loss can be estimated by the following expression [27],

$$\phi_{support} = \beta\left(\frac{t}{L}\right)^3,\qquad(10)$$

Typical values for the constant $\beta$ lie in the range 2 to 3 [27,28]. This would give a limiting loss factor of between $1.4 \times 10^{-9}$ and $2 \times 10^{-9}$. This is significantly below the measured losses for this particular cantilever. However 'stick-slip' losses may also exist associated with friction at the clamped end of the oscillating sample [13].

## E. Other Losses

As previously discussed, for both modes measured there is clear evidence of excess dissipation above that estimated for the sources detailed in section 4 through parts A to D.

An intermittent dissipation peak was seen in the measured loss of the fifth mode of the cantilever of length 57 mm. This peak was most likely due to energy loss into the clamping structure. To investigate this, a piezo transducer was attached to the upper part of the clamp to sense displacements of the clamp which could result from energy coupling to the clamp from an excited mode of the cantilever. This was compared to the estimated excess loss measured in this cantilever. The excess loss was found by subtracting the calculated thermoelastic loss from the measured total loss. In the following plots, each mode was excited to a similar amplitude and the magnitude of the peak at the relevant modal frequency was found using a spectrum analyser. The magnitude of the signal sensed by the piezo was then divided by the amplitude of the signal from the oscillating cantilever to normalise the piezo data.

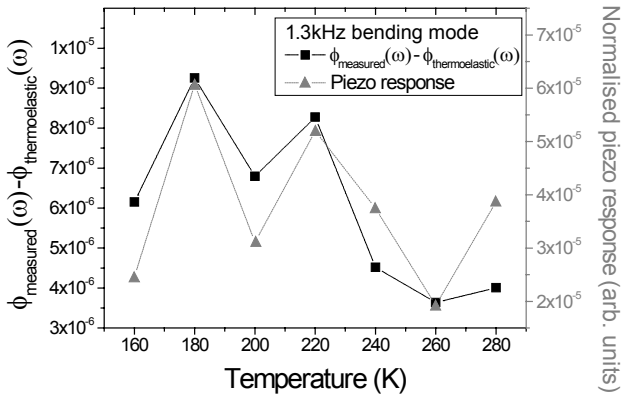

Fig. 8. Plot of the average excess loss of the fourth bending mode at $f \approx 1.3$kHz compared with the normalised magnitude of the signal from the piezo sensor.

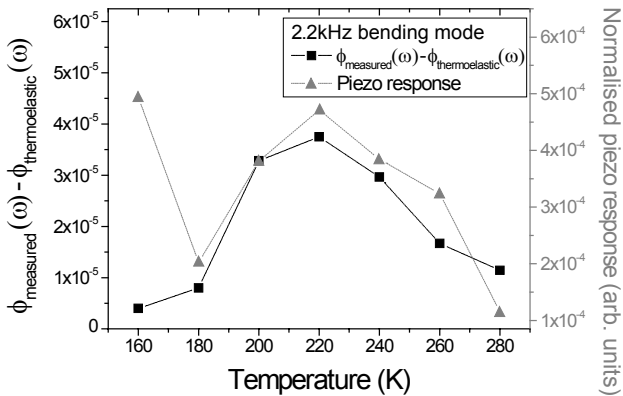

Fig. 9. Plot of the average excess loss of the fifth bending mode at $f \approx 2.2$kHz compared with the normalised magnitude of the signal from the piezo sensor.

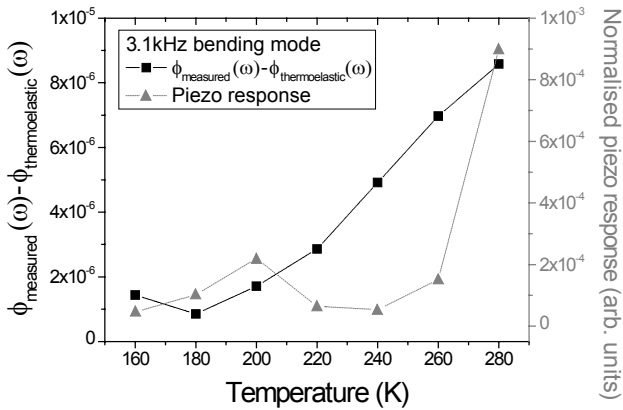

Fig. 10. Plot of the average excess loss of the sixth bending mode at $f \approx 3.1$kHz compared with the normalised magnitude of the signal from the piezo sensor.

In general the above plots indicate a clear relationship between the excess loss factors measured in this sample and the energy coupling to the clamp at the same frequency. Further work will be carried out to attempt to damp the resonances in the clamping structure or to shift the resonances by adding mass. In particular F.E. analysis has identified resonances associated with the flexing of the upper part of the clamp to be possible candidates for the parasitic resonances seen. Using a larger upper clamping block may reduce this effect.

## V. CONCLUSIONS AND FUTURE WORK

Our measurements of the mechanical dissipation of single crystal silicon cantilevers as a function of temperature in general show the dissipation decreasing as temperature decreases. At room temperature the measured dissipation is strongly dependent on the level of thermoelastic dissipation in the sample, however at lower temperatures other loss mechanisms become dominant. Losses associated with the surface of the samples are expected to be significant, but at a level lower than the measured losses. The level of surface loss will be investigated further by measuring the dissipation in samples of different thicknesses. A possible source of loss requiring further investigation is that of frictional losses associated with the end of the sample moving inside the clamp. To reduce this effect, samples with a greater ratio of thickness of clamp end to thickness of cantilever will be studied. Additionally, work will be focused towards reducing the energy coupling from the cantilevers into resonances within the clamping structure by modifying the clamp design.

In particular we do not see any distinct peaks in the dissipation close to 125K [9], at the level of dissipation found here. In common with other researchers, we have observed intermittent dissipation peaks at other temperatures in several of our experimental runs. However, we believe these are not intrinsic to our silicon samples but rather here are due to couplings to the clamping structure used.

### A. Acknowledgments

We would like to thank our colleagues in the GEO 600 project for their interest. We wish to thank V. Mitrofanov and J. Faller for useful discussions. We are grateful for the financial support provided by PPARC and the University of Glasgow in the UK and the NSF in the USA (award no: NSF PHY-0140297).

### B. References

[1] V.B. Braginsky, V.P. Mitrofanov, V.I. Panov, *Systems with small dissipation*, University of Chicago Press (1985).
[2] T. Uchiyama, D. Tatsumi, T. Tomaru, M.E. Tobar, K. Kuroda, T. Suzuki, N. Sato, A. Yamamoto, T. Haruyma, T. Shintomi, Phys. Lett. A 242 (1998) 211.
[3] K. Kuroda, Int. J. Mod. Phy. D 557 (1999) 8.
[4] S. Rowan, Proceedings of SPIE, 4856 (2003) 292.
[5] A. Giazotto, G. Cella, Class. Quant. Grav. 21, (2004) S1183.
[6] W. Winkler, K Danzmann, A. Ruediger, R Schilling, Phys. Rev. A 7022 (1991).
[7] V.B. Braginsky, M.L. Gorodetsky, S.P. Vyatchanin, Phys. Lett. A 264 (1999) 1.
[8] K. Numata, G.B. Bianc, M. Tanaka, S. Otsuka, K. Kawabe, M. Ando, K. Tsubono, Phys. Lett. A 284 (2001) 162.
[9] Y.S. Touloukian, E.H. Buyco, Plenum, New York (1970).
[10] G. Benning, C.F. Quate, C. Gerber, Phys. Rev. Lett. 56 (1986) 930.
[11] T.D.Stowe, K. Yasumura, T.W. Kenny, D. Botkin, K. Wago, D. Rugar, Science 264 (1994) 1560.

[12] K.Y. Yasumura, T.D. Stowe, E.M. Chow, T. Pfafman, T.W. Kenny, B.C. Stipe, D. Rugar, IEEE Trans. Electron Devices 9 (2000) 119.

[13] T.J. Quinn, C.C. Speake, R. S. Davis, W. Tew, Phys. Lett. A, 197 (1995) 197.

[14] P. Saulson, *Fundamentals of Interferometric Gravitational Wave Detectors*, World Scientific (1994).

[15] S.M. Twyford, PhD thesis, University of Glasgow (1998).

[16] J.B. Wachtman, W.E. Tefft, D.G. Lam, C.S. Apstein, Phys. Rev. 122 (1961) 1754.

[17] N.W. McLachlan, *Theory of Vibrations*, Dover Publications (1951) 96.

[18] U. Gysin, S. Rast, P. Ruff, E. Meyer, Phys. Rev. Lett B 69 (2004) 045403-2.

[19] L. Boernstein, Band 17 Halbleiter, Springer-Verlag Berlin, (1982).

[20] R. Lifshitz, M.L. Roukes, Phys. Rev. Lett. B 61 (2000) 5600.

[21] J. Yang, T. Ono, M. Esashi, IEEE Trans. Electron Devices 11 (2002) 779.

[22] Y. Wang, J.A. Henry, A.T. Zehnder, M.A. Hines, J. Phys. Chem. B 107 (2003) 14270.

[23] P. Mohanty, D.A. Harrington, K.L. Ekinci, Y.T. Yang, M.J. Murphy, M.L. Roukes, Phys. Rev. B 66 (2002) 085416.

[24] C.C. Lam, PhD Thesis, University of Rochester USA (1979).

[25] D.F. McGuigan, C.C. Lam, R.Q. Gram, A.W. Hoffman, D.H. Douglass, H.W. Gutche, J. Low Temp. Phys. 30 (1978) 621.

[26] V. Mitrofanov, Private Communication (2004).

[27] H. Hosaka, K. Itao, K. Kuroda, Sensors and Actuators A (1995) 87.

[28] K.Y. Yasumura, PhD Thesis, Stanford University USA (2001).